\documentclass[twocolumn,letterpaper]{IEEEAerospaceCLS}  

\usepackage{url}
\usepackage[]{graphicx} 
\usepackage{comment}
\newcommand{\ignore}[1]{}  
\newcommand{\degree}{^\circ}
\usepackage{epstopdf}

\begin{document}
\title{Impact of Lunar Dust on Free Space Optical (FSO) Energy Harvesting}

\author{%
Mohamed Naqbi\\ 
Polytechnique Montréal\\
Ecole Polytechnique de Bruxelles\\
2500 Chem. de Polytechnique\\
Montréal, Qc, H3T1J4\\
mohamed.naqbi@polytml.ca
\and 
Sébastien Loranger\\
Polytechnique Montréal\\
2500 Chem. de Polytechnique\\
Montréal, Qc, H3T1J4\\
sebastien.loranger@polymtl.ca
\and
Gunes Karabulut Kurt\\
Polytechnique Montréal\\
2500 Chem. de Polytechnique\\
Montréal, Qc, H3T1J4\\
gunes.kurt@polytml.ca
\thanks{\footnotesize 979-8-3503-0462-6/24/$\$31.00$ \copyright2024 IEEE}              
}

\maketitle

\thispagestyle{plain}
\pagestyle{plain}

\maketitle

\thispagestyle{plain}
\pagestyle{plain}

\begin{abstract}
In the backdrop of escalating ambitions for space exploration, particularly with programs like Artemis aiming for a sustainable human presence on the Moon, the issue of efficient and reliable energy transmission has become a critical concern. This paper focuses on the impact of levitating lunar dust particle above the lunar surface on energy harvesting through a Free Space Optics (FSO) beam, commonly known as \textquoteleft laser power beaming \textquoteright . This study employs known mathematical models and numerical analysis to examine a ground-to-ground link on the lunar surface, using realistic data from the literature. More precisely, it assesses transmission losses and the performance of FSO energy harvesting in the presence of lunar dust under both illuminated and dark lighting conditions. The findings reveal that the choices regarding the height above the lunar surface and the transmission distance are key determinants in evaluating both losses and energy harvesting efficiency. Specifically, it was observed that the transmit power required to operate a 250W Polaris rover at a distance of 20km was on the order of several kilowatts. Additionally, it was noted that in the Moon's darker regions, the impact of lunar dust on transmission was nearly negligible, making these areas more conducive for FSO-based energy transmission.
\end{abstract}

\tableofcontents

\section{Introduction}

In light of the escalating ambitions of space endeavors, exemplified by programs such as Artemis that aim to establish a sustainable human presence on the Moon, the issue of surface energy transmission has become increasingly critical. The growing focus on wireless power transmission techniques for lunar applications is attributable to the inherent challenges associated with deploying cabling on the Moon's intricate and inhospitable terrain. Specifically, transporting substantial quantities of electrical cables to the Moon for the purpose of establishing energy connections between various lunar ground stations would not only incur significant weight and volume costs but also present considerable implementation complexities. Each kilogram of material conveyed to the Moon necessitates a substantial amount of fuel for both launch and landing phases, thereby markedly elevating the overall mission expenditures. Furthermore, the installation of such cables on the Moon's rugged and dust-laden terrain would introduce additional operational challenges, particularly in terms of maintenance and durability due in part to the abrasiveness of lunar dust \cite{marcinkowski_lunar_2023}.\\

The Artemis mission is targeting the lunar South Pole for its landing sites. This region is of particular interest due to the presence of  peaks of eternal light (PELs), which receive almost continuous sunlight and permanently shadowed regions (PSRs), which are potential sites for ressources such as water ice. These contrasting conditions are ideal for the application of laser power beaming technology, which could provide a continuous power supply in shadowed areas by transmitting energy wirelessly from illuminated regions \cite{Grandidier2021LaserPower}.

Various technological systems utilize microwave radiofrequency (RF) for the purpose of wireless power transmission (WPT), offering distinct advantages in aspects such as conversion efficiency and spatial coverage \cite{microSat2Earth,chen_novel_2019}. Nevertheless, this modality is accompanied by inherent limitations, such as vulnerability to electromagnetic interference. In juxtaposition, free-space optical (FSO) technology leverages collimated laser diodes to concentrate energy transmission within a more circumscribed circular region \cite{he_analysis_2021}. Consequently, the energy received by a targeted photovoltaic cell area remains undiminished in FSO-based WPT, in contrast to the attenuation observed in RF-based systems \cite{donmez_mitigation_2023}.\\

FSO systems offer notable advantages in efficiency and targeting precision. However, these systems face unique challenges when deployed in lunar environments. One such challenge is beam misalignment, which can occur due to the Moon's irregular terrain or mechanical vibrations, thereby affecting the laser beam's accuracy \cite{alouiniEH2}. Another potential issue is the presence of electrostatically charged lunar dust particles. These charged particles can stick to surfaces, potentially impairing the operation of instruments, obstructing optical lenses, and interfering with mechanical and electrical systems \cite{stubbs2007impact}. In addition, these particles can levitate above the lunar surface due to the 'dusty plasma' phenomenon and have the potential to scatter or absorb the laser light \cite{popel_dusty_2015}. However, this phenomena is much less prevalent in regions of the Moon not exposed to solar illumination, such as shadowed craters or polar areas. In these locations, the absence of solar light diminishes the likelihood of surface ionization, subsequently reducing both plasma formation and the electrostatic charging of dust particles \cite{popel_dusty_2013}.\\

Although prior studies have examined FSO energy harvesting \cite{jin_wireless_2019} and the complications introduced by lunar dust, a conspicuous research gap exists in the integration of these two elements. The present study seeks to bridge this knowledge void by incorporating a lunar dust attenuation model into an FSO energy harvesting framework. Constructed based on a corpus of empirical and theoretical data gleaned from extant scientific literature, this model offers a unified structure for comprehensive analyses.\\

In our paper, we specifically investigate the impact of levitating lunar dust, attributed to the 'dusty plasma' phenomenon, on FSO-WPT. Our study examines varying heights above the lunar surface, extending up to a maximum height of 2 meters and applies to a short-range application of maximum 100 km between a lunar transmitter and a receiving unit. Such a receiving unit could be, for instance, an exploratory rover deployed on the lunar surface.\\

The principal contributions of the present research are enumerated as follows:
\begin{itemize}
    \item The study employs Mie theory to model a lunar dust particle as a single stratified sphere.
    
    \item A modified version of the Beer-Lambert law is utilized to quantify the attenuation effects of the laser beam as it traverses a medium containing lunar dust particles.
    
    \item Empirical data concerning the size, density, and refractive index of lunar dust are meticulously collated from existing scientific literature, thereby ensuring the model's fidelity to real-world conditions.
    
    \item FSO energy harvesting model is constructed based on realistic parameters, enhancing the applicability and reliability of the study's findings.
\end{itemize}

The remainder of this paper is structured as follows. In Section II, we introduce the system model for FSO-based energy harvesting in a lunar environment, incorporating a realistic model to assess the impact of lunar dust under varying conditions. Section III elaborates on the simulation parameters and evaluates the efficacy of the FSO-based energy harvesting system in both lunar-illuminated and lunar-dark scenarios. The section also includes a discussion on the system's suitability for lunar rovers. Section IV concludes the paper and outlines future research directions.

\section{System model}


\begin{figure*}[h!]
    \centering
    \includegraphics[width=6in]{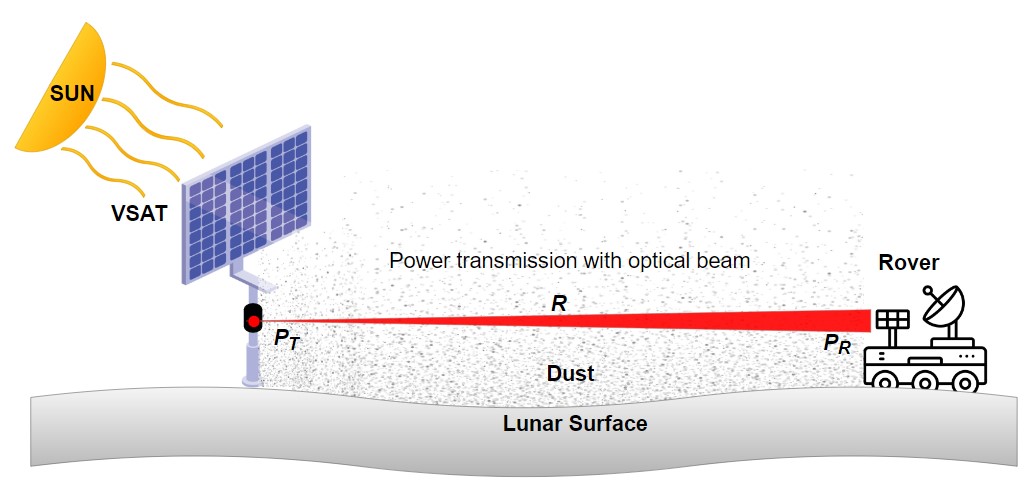}
    \caption{\bf{Lunar Surface Optical Power Transmission System}}
    \label{A}
\end{figure*}

Our proposed system model, as demonstrated in Figure \ref{A}, consists of a transmitter (e.g., lunar vertical solar array technology (VSAT) or nuclear fission reactors \cite{marcinkowski_lunar_2023}) and a receiver unit (e.g., Amalia or Polaris rovers \cite{kim_multidisciplinary_2020}) based on the lunar surface. In the proposed system, the receiver unit employs energy harvesting techniques from a remote laser diode featuring adaptive beam divergence, as delineated in \cite{adaptivebeamdivergence}. This mechanism ensures the maintenance of an optimal spot diameter as the separation between the transmitter and receiver increases. The system is designed to be self-sustaining and aims to utilize wavelength-dependent conversion efficiencies that are as high as feasible. It is imperative to note that the efficiency metrics must be consistent across the same wavelengths (\( \lambda \)) for both the transmitting and receiving units, as indicated in \cite{EHwindow}. Consequently, meticulous selection of system components is essential for optimal performance.

The harvested power of a free space LoS optical link is expressed as follows \cite{ComprehensivePathLoss}
{\footnotesize
\begin{equation}
P_{h} = P_{t} \left( \frac{\lambda}{4\pi R} \right)^{2} \eta_{e/o}(\lambda) \eta_{h}(\lambda) L_{t}(\psi_{t}) G_{t} L_{r}(\psi_{r}) G_{r} L_{e} L_{s} L_{c},
\end{equation}
}where \( P_{h} \) is the harvested electrical power, \( P_{t} \) is the transmitted electrical (input) power, \( \eta_{e/o}(\lambda) \) and \( \eta_{h}(\lambda) \) are the wavelength-dependent electrical-to-optical power conversion efficiency (PCE) and energy harvesting conversion efficiency (EHCE), respectively. In addition, \( L_{t}(\psi_{t}) \) is the radial angle-dependent misalignment loss factor at the transmitter, \( G_{t} \) is the transmitter gain, \( L_{r}(\psi_{r}) \) denotes the radial angle-dependent misalignment loss factor at the receiver, \( G_{r} \) is the receiver gain, \( L_{e} \) is the atmospheric extinction/attenuation loss, \( L_{s} \) is the scintillation loss, and \( L_{c} \) represents the fiber coupling loss.\\

For our lunar FSO energy harvesting scenario, we consider $L_{e} = L_{s} = L_{c} = 1 $, as done in \cite{ComprehensivePathLoss}. Moreover, in order to account for the influence of lunar dust on optical transmission, this study introduces the loss factor \( L_{\text{dust}} \), which quantifies the loss attributed to the scattering and absorption effects of lunar dust. Therefore, the received power of the free space lunar line of sight (LoS) optical link becomes 
{\footnotesize
\begin{equation}
P_{h} = P_{t} \left( \frac{\lambda}{4\pi R} \right)^{2} \eta_{e/o}(\lambda) \eta_{h}(\lambda) L_{t}(\psi_{t}) G_{t} L_{r}(\psi_{r}) G_{r} L_{dust}.
\end{equation}
}

The transmitter and receiver gains can be approximated as \cite{Shlomi_optimization_2004}
\begin{equation}
G_{t} \approx \left( \frac{\pi d_{t}}{\lambda} \right)^2,
\end{equation}
\begin{equation}
G_{r} \approx \left( \frac{\pi d_{r}}{\lambda} \right)^2,
\end{equation}
where \( d_{t} \) and \( d_{r} \) are the aperture diameter of the transmitter and receiver, respectively.\\

The loss factor due to misalignment at both the transmitter and receiver can be calculated as follows assuming Gaussian beam \cite{Shlomi_optimization_2004} 
\begin{equation}
    L_{t} = \exp(-G_{t}\psi_{t}^2),
\end{equation}
\begin{equation}
    L_{r} = \exp(-G_{r}\psi_{r}^2),
\end{equation}

where $\psi_{t}$ and $\psi_{r}$ are the transmitter’s and receiver’s radial misalignment error angle.\\

The model for the misalignment error angle is based on the methodology presented in \cite{donmez_mitigation_2023}.

\subsection{Lunar Dust Attenuation Model}

The presence of lunar dust adversely affects laser transmission through mechanisms of scattering and absorption. The magnitude of this impact is contingent upon variables such as the wavelength of the laser and specific lunar dust characteristics. This study introduces a comprehensive lunar dust attenuation model, built upon realistic data collected from existing literature. The model employs Mie theory to simulate the scattering and absorption characteristics of lunar dust particles \cite{craig_f_bohren_donald_r_huffman_absorption_1998}.
\\
\subsubsection{Mie theory}
 Lunar dust particles are angular and irregular in shape \cite{liu_characterization_2008}. However, for the application of Mie theory, a spherical approximation is adopted. This approximation is particularly useful as it significantly simplifies the equations and associated calculations, making the analysis more manageable. Although this simplification may not capture all the complex details of the particles' actual shapes, it provides a good initial approach for understanding their absorption and scattering properties.\\

In Mie theory, the scattering and absorption properties of a spherical particle are described by a set of complex mathematical equations that take into account the size and refractive index of the particle, as well as the wavelength of the incident light. Unlike Rayleigh scattering, which is applicable for particles much smaller than the wavelength of light, Mie theory is applicable for particles of any size, including those comparable to or larger than the wavelength. The theory provides a way to calculate the scattering and absorption cross-sections, which are critical for understanding how much light is scattered or absorbed by the particle. Due to the complexity of these equations, numerical methods are often employed to find solutions, as they can be challenging to solve analytically \cite{wriedt_mie_2012}.


The Mie theory is used to calculate the extinction cross-section $C_{ext}$ as \cite{craig_f_bohren_donald_r_huffman_absorption_1998}\begin{equation}
C_{ext} = \frac{2\pi}{k^2} \sum_{n=1}^{\infty} (2n + 1) Re(a_n + b_n),
\end{equation}
\begin{equation}
C_{ext} = C_{abs} + C_{scat}
\end{equation}
In this context, \( C_{\text{ext}} \) represents the sum of the absorption and scattering cross-sections, \( k = \frac{2\pi}{\lambda} n_m \) denotes the wavenumber in the medium, and \( a_n \) and \( b_n \) are the expansion coefficients.

The expansion coefficients \( a_n \) and \( b_n \) are defined as follows \cite{craig_f_bohren_donald_r_huffman_absorption_1998}

\begin{equation}
a_n = \frac{m \psi_n(m x) \psi_n'(x) - \psi_n(x) \psi_n'(m x)}{m \psi_n(m x) \xi_n'(x) - \xi_n(x) \psi_n'(m x)},
\end{equation}

\begin{equation}
b_n = \frac{\psi_n(m x) \psi_n'(x) - m \psi_n(x) \psi_n'(m x)}{\psi_n(m x) \xi_n'(x) - m \xi_n(x) \psi_n'(m x)}
\end{equation}

In this setting, \( \psi_n(x) \) and \( \psi_n'(x) \) refer to the Riccati-Bessel function of the first kind and its derivative, respectively. Similarly, \( \xi_n(x) \) and \( \xi_n'(x) \) denote the Riccati-Bessel function of the third kind (also known as the spherical Hankel function of the first kind) and its derivative, respectively. Additionally, \( m \) represents the relative refractive index of the sphere to the surrounding medium, and \( x = k r \) serves as the size parameter, where \( k \) is the wavenumber in the medium and \( r \) is the radius of the sphere.

$\psi_n(x)$ and $\xi_n(x)$ are calculated as follows \cite{craig_f_bohren_donald_r_huffman_absorption_1998}

\begin{equation}
\psi_n(x) = \sqrt{\frac{\pi x}{2}} J_{n + \frac{1}{2}}(x),
\end{equation}

\begin{equation}
\xi_n(x) = \sqrt{\frac{\pi x}{2}} H_{n + \frac{1}{2}}^{(1)}(x)
\end{equation}
where \( J_{n + \frac{1}{2}}(x) \) represents the Bessel function of the first kind, while \( H_{n + \frac{1}{2}}^{(1)}(x) \) denotes the Hankel function of the first kind.

\subsubsection{Beer-Lambert law generalisation}

In a lunar setting, FSO technology employs laser beams to enable wireless power transmission across varying distances. While terrestrial applications of FSO technology are often hindered by atmospheric conditions such as fog, rain, or snow \cite{yahia_haps_2022}, the lunar environment introduces its own challenge in the form of levitating dust particles, attributed to the 'dusty plasma' phenomenon. For subsequent considerations, it is important to note that the particle density is assumed to be uniform along the transmission path in both the sunlit and dark regions of the Moon.\\

FSO employs a laser beam with a circular radius \(a\) at the point of emission, which propagates over a distance \(l\) in the vacuum of lunar space. The volume covered by the light forms a truncated cone with an angle \(\theta\). The received beam's radius is given by \cite{lacaze_gaps_2009}
\[
b = a + l \tan \theta
\]
where \(a\) is the initial radius of the beam.

Therefore, in FSO systems, lasers emit a light beam that propagates in the shape of a truncated cone rather than in a straight line. This unique characteristic means that the intensity of the beam does not strictly follow the Beer-Lambert law, which is commonly used to describe light absorption in a medium. However, according to \cite{lacaze_gaps_2009}, in a homogeneous environment, the deviation from this law is generally very slight. Therefore, the use of the Beer-Lambert law remains a justified approximation for assessing the intensity of the laser beam in FSO systems \cite{lacaze_gaps_2009}.

Traditionally, the Beer-Lambert law is formulated in terms of optical depth, but it can also be expressed through an attenuation coefficient, \( \alpha \), which is an integrated measure of absorption and scattering in the medium \cite{lacaze_gaps_2009}. This coefficient is defined as 
\begin{equation}
\alpha = N \times C_{ext},
\end{equation}
where \( N \) is the particle density and \( C_{ext} \) is the total extinction cross-section. The transmittance of the system, define as a loss factor $L_{dust}$ for a path length \( L \) is then given by \cite{lacaze_gaps_2009}
\begin{equation}
L_{dust} = e^{-\alpha L}.
\end{equation}
This formalism offers a unique metric, \( \alpha \), that encapsulates contributions from both absorption and scattering, thus allowing for a simplified yet comprehensive evaluation of the system's loss factor $L_{dust}$. 

\subsubsection{Lunar dust application}

In order to apply this model, data specific to the properties of lunar dust—such as particle size distribution, density, and refractive index—are required. The grain-size distributions of lunar dust from the Apollo 11 and 17 missions were analyzed using SEM techniques. The particle size distributions ranged from 20 $\mu$m to 20 nm and followed a log-normal pattern, with peak modes between 100 and 300 nm \cite{park_characterization_2008,vidwans_size_2022}.

In this study, lunar dust particle sizes were quantified using equivalent circular diameters, calculated as \(2 \sqrt{\frac{\text{Particle Surface Area}}{\pi}}\). This metric facilitates a shape-independent comparison of particle sizes.

Concerning the particle density, recent studies have focused on the dusty plasma systems near the Moon's surface, specifically addressing the levitation and charging of lunar dust particles \cite{popel_dusty_2015,popel_dusty_2013,popel2018lunar}. These particles become charged due to the influence of photoelectrons, solar wind electrons, and ions, as well as solar radiation. \\

\begin{figure}[h!]
    \centering
    \includegraphics[width=3.25in]{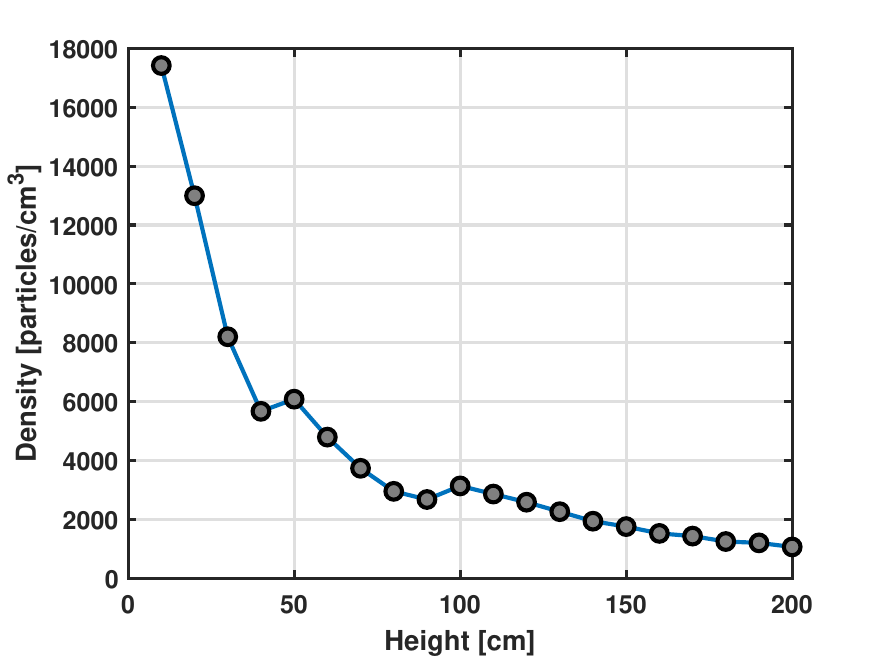}\\
    \caption{\textbf{Lunar dust particle density as a function of the near-surface height over the illuminated part of the Moon from \protect\cite{popel_dusty_2015}}}
    \label{Fig2}
\end{figure}

In our effort to determine the distribution of the dust grains at various heights above the lunar surface, the results presented in \cite{popel_dusty_2015} are taken. According to this source, the number density of charged dust grains in the near-surface layer over the illuminated part of the Moon is on the order of $10^3 \, \text{cm}^{-3}$. This high density is attributed to the large amount of photoelectrons, including those emitted from the surfaces of flying dust grains. The distribution of density of lunar dust particles by near-surface height in illuminated areas is shown in
Figure~\ref{Fig2}, which is reproduced from \cite{popel_dusty_2015}.

The distribution is determined using reference parameters specified in \cite{popel_dusty_2015}: the subsolar angle is \( \theta = 77^\circ \), the density of photoelectrons emitted immediately at the lunar surface is \( N_0 = 1.9 \times 10^5 \, \text{cm}^{-3} \), and the thermal energy of the electrons is \( T_e = 0.1 \, \text{eV} \). A rather large value of $\theta$ was taken because the corresponding latitude region of interest is the polar regions of the Moon.

In a plasma over the dark part of the Moon the photoelectrons are expected to be absent \cite{popel2018lunar}. The lunar surface is charged negatively under the influence of the solar wind electrons and the plasma of the terrestrial magnetosphere tail. Therefore, in those regions, the dust particle density is significantly lower, ranging between $10^{-2}$ and $10^{-1}$ $cm^{-3}$ for dust particles of sizes \( a \approx 100 \, \text{nm} \) \cite{popel_dusty_2015}. Therefore, for the dark regions of the Moon, we postulate that the particle density follows the same distribution as in illuminated areas, albeit scaled by a factor of \(10^{-4}\).
\\

Regarding the lunar dust complex refraction index 
\begin{equation}
    \tilde{n} = n + ik,
\end{equation}
the real component \( n \) serves to describe the phase velocity of light as it propagates through the medium, which in this study is represented by the lunar dust particle. This real part plays a key role in governing optical phenomena such as refraction. On the other hand, the imaginary component \( k \) is responsible for quantifying the absorption of electromagnetic waves during their passage through the medium.\\

In the existing literature, the real component of the complex refractive index for lunar dust typically ranges between 1.58 and 1.78 \cite{refractiveindex1,refractiveindex2}. Obtaining precise values for the imaginary component presents a greater challenge. For the purposes of this study, we approximate the refractive index by considering lunar dust particles to be rich in silicates. Consequently, we reference 'astronomical silicate' as the basis for our refractive index values \cite{draine_optical_1984}. This approximation is employed to simplify the modeling process, given the inherent complexity of lunar dust due to its variable composition. Geological studies of the Moon have revealed that the lunar regolith is rich in silicate minerals. Furthermore, astronomical silicates are well-characterized in terms of their optical properties, providing a robust basis for this simplification.\\

Utilizing the dielectric function data $(\epsilon_{1},\epsilon_{2})$ as outlined in the literature \cite{draine_optical_1984}, the complex refractive index is determined through the subsequent equations

\begin{equation}
n = \sqrt{\frac{\sqrt{\epsilon_1^2 + \epsilon_2^2} + \epsilon_1}{2}},
\label{eq:A1}
\end{equation}
\begin{equation}
k = \sqrt{\frac{\sqrt{\epsilon_1^2 + \epsilon_2^2} - \epsilon_1}{2}}.
\label{eq:A2}
\end{equation}

\subsection{Energy Harvesting Model}

In the context of lunar surface links, the LoS distance range under consideration spans from 50 m to 100 km. FSO energy harvesting emerges as a supplementary energy source, and is particularly relevant when solar energy cannot be harvested, as is the case on long lunar nights or in permanently shadowed region. The transmitted energy is harvested using solar cells which are optimally designed to convert frequencies within the solar spectrum. Given that laser diodes predominantly operate in the infrared band, specifically at 1064 nm, solar cells are well-suited for FSO based energy harvesting \cite{Solarcells}. Among the various types of solar cells capable of converting optical energy to electrical energy, InGaAsP-based solar cells, which offer a conversion efficiency of 26.4\%, are compatible with our proposed system operating at a laser wavelength of 1064 nm \cite{donmez_mitigation_2023,algora_beaming_2022}.

\section{Attenuation and Performance Evaluation}

The impact of lunar dust on the system's performance is quantitatively assessed through simulation studies. Prior to calculating the average harvested power, the cross-sectional area of the lunar dust particles under consideration is determined. Subsequently, the attenuation coefficient and the loss factor \( L_{\text{dust}} \) are evaluated for two distinct lunar scenarios: the illuminated and the dark regions. The average harvested power is then computed for both scenarios. Statistical analyses of independent elevation and azimuth misalignment error angles for both the receiver and transmitter are conducted using the Monte Carlo method, with pointing resolution parameters outlined in Table \ref{tab:parameters}.

\subsection{Simulation Parameters}

As delineated in subsection 'Lunar Dust Application' the grain size of lunar dust adheres to a log-normal distribution with peak modes ranging between 100 and 300 nm. Upon an analysis of the distribution \cite{park_characterization_2008}, this study opts for an average effective diameter of 150 nm for the simulation.

The judicious selection of the laser diode type is imperative for the system's efficacy. The operating wavelength must not only facilitate adequate energy harvesting but also be compatible with space mission requirements. In light of these considerations and the findings of \cite{lasertypes}, we employ a Yd: NVO4 1064 nm laser source with a PCE of 51\% \cite{laser1064}.

Based on the dielectric function data provided in \cite{draine_optical_1984} and the formulas \ref{eq:A1} and \ref{eq:A2}, the complex refractive index for a wavelength of 1064 nm is determined to be \( n = 1.733 + i 0.05 \). The value of the real part of the refractive index is within the range of values given in the literature for lunar dust, which solidifies the relevance of the chosen approach. Given the lunar conditions, the refractive index of the medium surrounding the particle is set to 1.0. In addition, the Mie scattering coefficients are calculated over an angular range of 180 $\degree$.

The adaptive beam divergence angle varies in relation to the spot diameter and distance using the small-angle approximation \cite{spotdiameter}

\begin{equation}
\theta \text{ [rad]} = \frac{\text{Spot Diameter [m]}}{R \text{ [m]}}.
\end{equation}

In addition, the appropriate transmitter aperture diameter, \( d_t \), can be determined by \( d_t \approx \frac{1.22\lambda}{\theta} \) \ based on the Fraunhofer diffraction.\\
In our study, we adjust the beam divergence angle to guarantee that the diameter of the spot at the receiver perfectly matches the dimensions of the receiving unit. Consequently, we select the diameter of the transmitter aperture to align with the necessary divergence angle. Furthermore, this flexibility allows us to consistently meet the criteria for Fraunhofer diffraction in our simulations. Finally, various transmit powers \( P_{t} \) between 1 W and 10 kW are considered in our simulations \cite{marcinkowski_lunar_2023}.

\begin{table}[h!]
    \centering
    \caption{Summary of Parameters}
    \label{tab:parameters}
    \begin{tabular}{|p{4cm}|c|}
        \hline
        \bf{Parameter} & \multicolumn{1}{c|}{\bf{Value}} \\
        \hline
        Particle Average Effective Diameter & 150 nm \\
        \hline
        Particle Complex Refractive Index & \(1.733 + i 0.05 \) \\
        \hline
        Medium Refractive Index & 1.0 \\
        \hline
        Angular Range & 180 $\degree$ \\
        \hline
        Laser Wavelength $(\lambda)$ & 1064 nm \cite{Shlomi_optimization_2004}\\
        \hline
        Laser Diode PCE (\%) & 51\% \cite{laser1064} \\
        \hline
        Beam Divergence Angle ($\theta$) & Adaptive \\
        \hline
        Polaris Rover Receiver Aperture Diameter (\(d_r\)) & 2.1 m \\
        \hline
        Pointing Resolution & 2 µrad \cite{Grandidier2021LaserPower} \\
        \hline
        EHCE (\%) & 26.4\% \cite{EHCE} \\
        \hline
        Polaris Rover Power Requirement (W) & 250 W \cite{kim_multidisciplinary_2020} \\
        \hline
    \end{tabular}
\end{table}

As previously stated, InGaAsP solar cells offer an EHCE of 26.4\%. Regarding the receiver aperture diameter, it is determined based on the area of the Polaris rover, which is \(3.6 \, m^2\), as cited in \cite{kim_multidisciplinary_2020}. This results in an aperture receiver diameter of approximately \(2.1 \, \text{meters}\). Notably, the energy requirement for this rover is \(250 \, \text{W}\).

\subsection{Results and Discussions}

\begin{figure}[h!]
    \centering
    \includegraphics[width=3.4in]{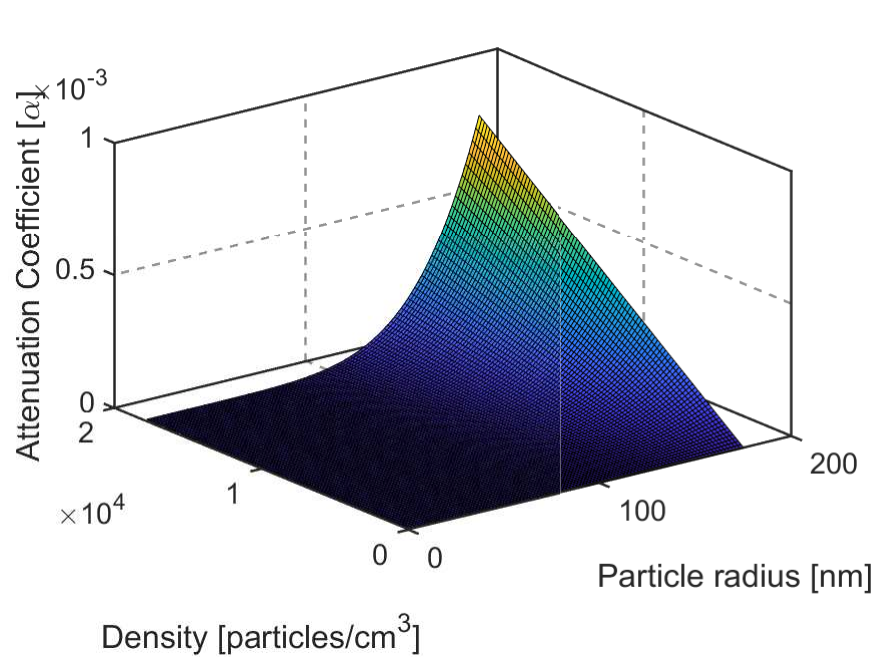}\\
    \caption{\textbf{Attenuation coefficient as a function of density and particle radius}}
    \label{Fig3}
\end{figure}

Simulations were carried out to assess the influence of lunar dust on the average harvested power \( P_{h} \) by the Polaris rover, considering varying transmit powers and distances \( R \) within our proposed system model. Additionally, the study identifies the minimum transmit power necessary to fulfill the energy requirements of the rover at varying distances from the transmitting source.

Firstly, it should be noted that an effective mean diameter of 150 nm was selected for simulations concerning harvested power. However, the attenuation coefficient can vary significantly depending on the chosen diameter value. Specifically, it appears to vary exponentially as the particle radius increases, as can be observed in Figure \ref{Fig3}. The density, on the other hand, will not be considered constant throughout the simulations; rather, it will vary with height as described in Figure \ref{Fig2}. This will have a linear impact on the attenuation coefficient, as shown in Figure \ref{Fig3}.

\begin{figure}[h!]
    \centering
    \includegraphics[width=3.25in]{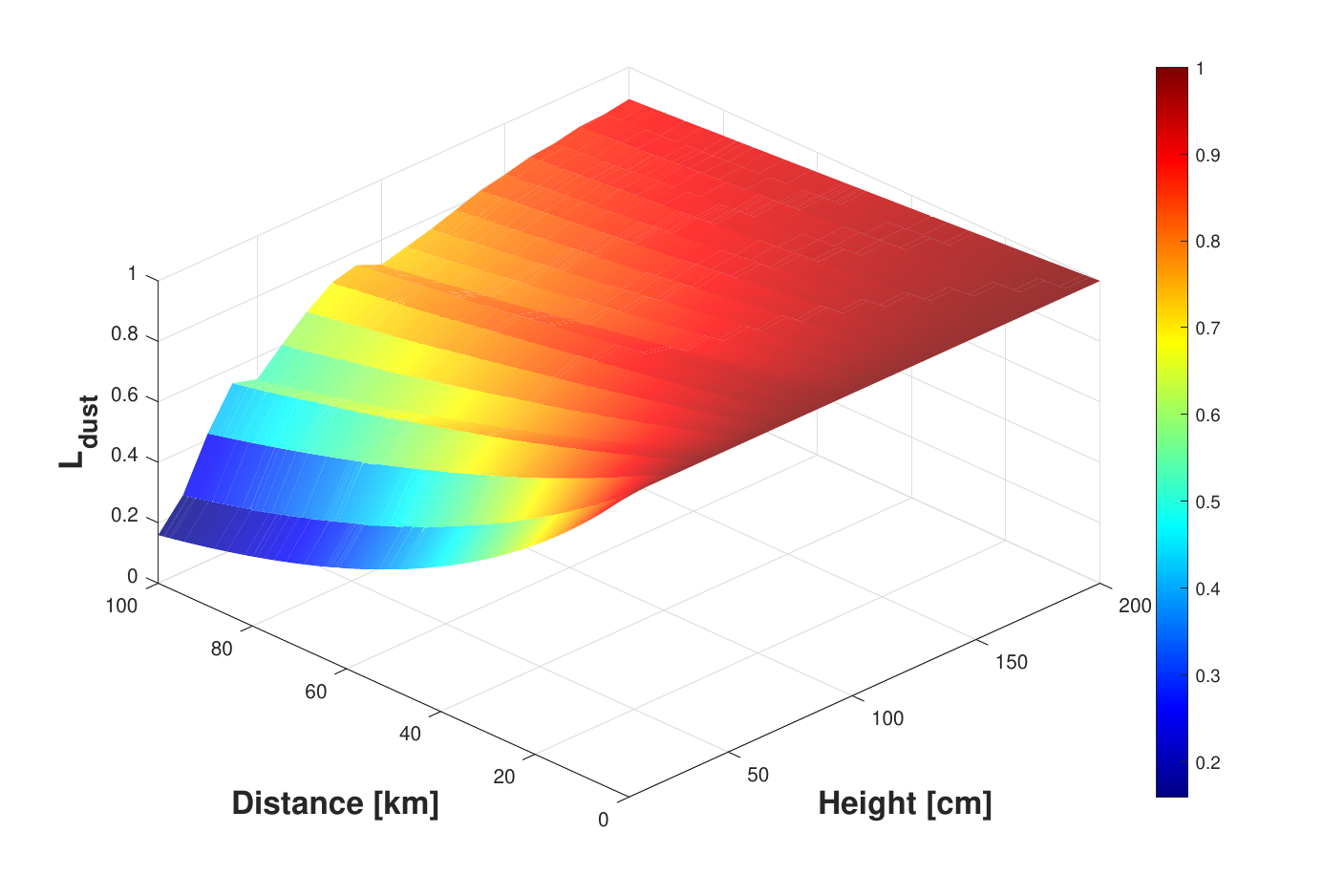}\\
    \caption{\textbf{\bf{$L_{dust}$ as a function of the transmission distance and the height above the lunar surface when the Moon is illuminated}}}
    \label{Fig4}
\end{figure}

\begin{figure}[h!]
    \centering
    \includegraphics[width=3.5in]{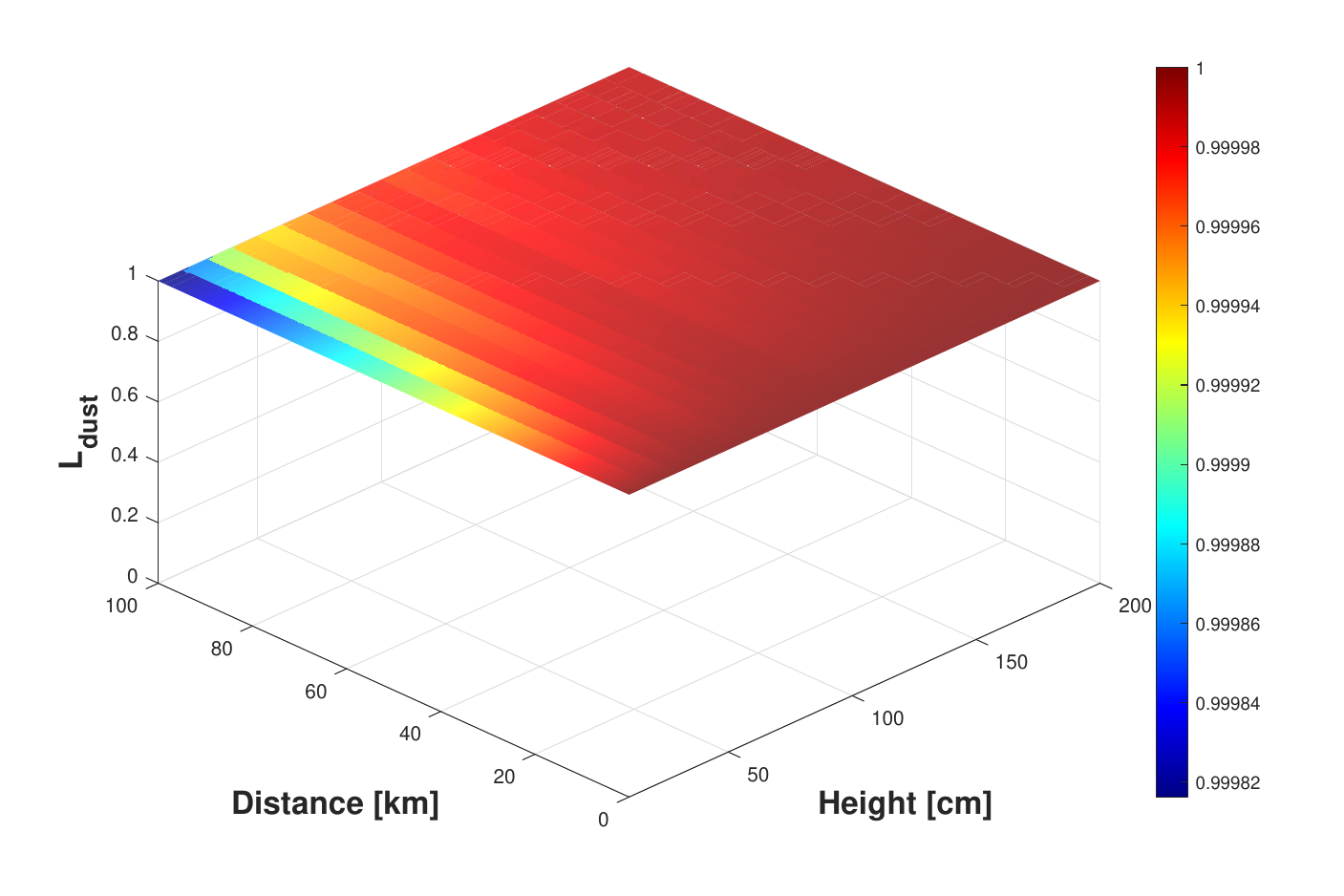}\\
    \caption{\textbf{$L_{dust}$ as a function of the transmission distance and the height above the lunar surface when the Moon is not illuminated}}
    \label{Fig5}
\end{figure}

The dust loss factor, which quantifies the laser transmittance through a medium partially composed of lunar dust, was determined through simulations. These simulations employed the dust attenuation model outlined in Section 2, along with the data specified in the sub-sections under 'Simulation Parameters.' This factor is presented as a function of both the transmission distance and the height above the lunar surface, for scenarios where the Moon is illuminated and where it is in darkness. The results are depicted in Figures \ref{Fig4} and \ref{Fig5}.\\

In the scenario where the Moon is illuminated, it is observed that as the height above the lunar surface increases, the dust loss factor \( L_{\text{dust}} \), which quantifies the losses attributed to dust, similarly increases. We note that this loss is most significant below a height of 1 meter. This might suggest that installing the devices at a height greater than 1 meter could greatly improve the system's overall efficiency. Conversely, as the transmission distance increases, \( L_{\text{dust}} \) decreases, a result that is both logical and expected. However, this relationship is not simply linear. Figure \ref{Fig4} reveals that below a certain critical height and beyond a specific critical distance, \( L_{\text{dust}} \) can fall below 0.5, leading to critical losses in the transmission.\\

In the case where the Moon is in darkness, which is the case in permanently shaded regions or on moonlit nights, the situation differs as it can be observed in Figure \ref{Fig5}. The variations of \( L_{\text{dust}} \) with respect to height and distance are minimal. Consequently, transmissions are only marginally affected by lunar dust. \\

Using the data presented in Table~\ref{tab:parameters}, the results of the average harvested energy as a function of transmitted power for a power range from \(1\, \text{W}\) to \(10\, \text{kW}\) and distance from \(50\, \text{m}\) to \(100\, \text{km}\) are shown in the figure. In the case where the Moon is illuminated, we can observe significant variations depending on the chosen transmitted power and the transmission distance. Logically, the harvested power is greater over the considered ranges when the height above the lunar surface increases and when the transmission distance decreases. \\

Considering the power requirements of the Polaris rover, which amount to \(250\, \text{W}\), and taking into account a transmission distance of \(20\, \text{km}\)~\cite{kim_multidisciplinary_2020}, we find that a minimum transmission power of \(4210\, \text{W}\) is required for a height of \(20\, \text{cm}\) above the ground, \(3480\, \text{W}\) for a height of \(50\, \text{cm}\), and \(2760\, \text{W}\) for a height of \(100\, \text{cm}\), to meet the rover needs. These results are perfectly consistent with the fact that at increased heights, the \(L_{\text{dust}}\) coefficient has a smaller influence, meaning that lunar dust has less effect on the transmission.
\begin{figure}[h!]
    \centering
    \includegraphics[width=3.5in]{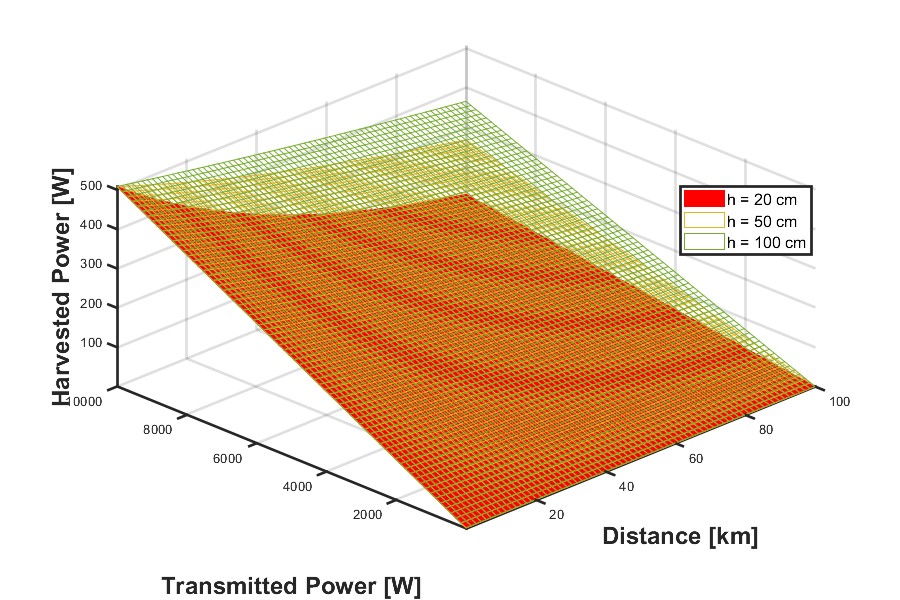}\\
    \caption{\textbf{Average harvested power for various heights when the Moon is illuminated}}
    \label{Fig6}
\end{figure}

\begin{figure}[h!]
    \centering
    \includegraphics[width=3.5in]{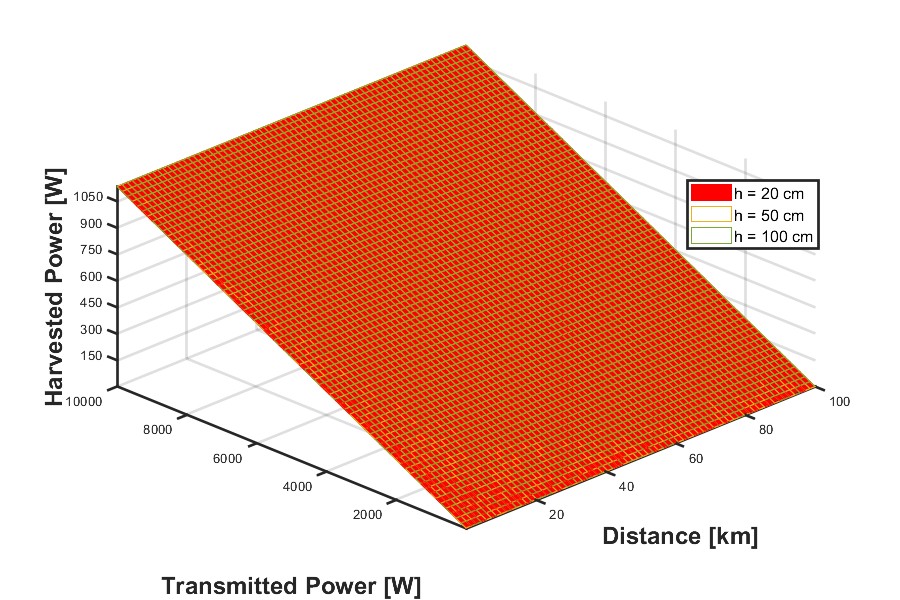}\\
    \caption{\textbf{Average harvested power for various heights when the Moon is not illuminated}}
    \label{Fig7}
\end{figure}

Regarding the scenario where the rover is in a dark region, Figure~\ref{Fig5} illustrates that the influence of lunar dust is negligible. Therefore, height no longer matters and the 3 planes coincide, as can be seen in Figure \ref{Fig7}. Considering the \(L_{\text{dust}}\) coefficient equal to 1, we find for a distance of \(20\, \text{km}\) that a transmission power of \(2050\, \text{W}\) is sufficient to meet the \(250\, \text{W}\) requirement of the rover. Once again, this result is consistent since, in this case, the effect of lunar dust is almost negligible compared to previous results.

\section{Conclusion}
This article has explored the impact of lunar dust on Free Space Optics (FSO) energy harvesting, also referred
to as \textquoteleft laser power beaming \textquoteright for short-range transmission. Utilizing known mathematical models and numerical simulations, we assessed transmission losses and the performance of FSO energy harvesting under various scenarios. Our findings indicate that the choice of the height above the lunar surface and the transmission distance are key factors in evaluating both losses and energy harvesting efficiency. Moreover, we observed that in darker regions of the Moon, the impact of lunar dust on transmission is negligible.

Although our study provides valuable insights, it does have certain limitations, particularly concerning the density of lunar dust, which can vary due to various environmental factors such as the subsolar angle. Future research could focus on more complex models that account for these variations.

The results of this study have significant implications for the design of energy transmission systems on the Moon, especially for transmissions in the illuminated part of the moon. A better understanding of the impact of lunar dust can lead to more efficient and reliable systems.  In summary, our research contributes to a better understanding of the challenges associated with lunar energy transmission and lays the foundation for more effective solutions to address these challenges.

\acknowledgments
This work was supported in part by the Tier 1 Canada Research Chair program. The author would like to acknowledger Dr. Irfan Azam and Baris Donmez for the insightful discussions.


\bibliographystyle{IEEEtran}
\bibliography{bibliography}

\begin{thebibliography}{10}
\providecommand{\url}[1]{#1}
\csname url@samestyle\endcsname
\providecommand{\newblock}{\relax}
\providecommand{\bibinfo}[2]{#2}
\providecommand{\BIBentrySTDinterwordspacing}{\spaceskip=0pt\relax}
\providecommand{\BIBentryALTinterwordstretchfactor}{4}
\providecommand{\BIBentryALTinterwordspacing}{\spaceskip=\fontdimen2\font plus
\BIBentryALTinterwordstretchfactor\fontdimen3\font minus \fontdimen4\font\relax}
\providecommand{\BIBforeignlanguage}[2]{{%
\expandafter\ifx\csname l@#1\endcsname\relax
\typeout{** WARNING: IEEEtran.bst: No hyphenation pattern has been}%
\typeout{** loaded for the language `#1'. Using the pattern for}%
\typeout{** the default language instead.}%
\else
\language=\csname l@#1\endcsname
\fi
#2}}
\providecommand{\BIBdecl}{\relax}
\BIBdecl

\bibitem{marcinkowski_lunar_2023}
A.~Marcinkowski, L.~Carrio, S.~Hilliard, C.~Edwards, A.~Elhawary, D.~Clem, M.~Blood, L.~May, and T.~Cichan, ``Lunar surface power architecture concepts,'' in \emph{2023 {IEEE} Aerospace Conference}.\hskip 1em plus 0.5em minus 0.4em\relax {IEEE}, pp. 1--19.

\bibitem{Grandidier2021LaserPower}
J.~Grandidier, P.~Jaffe, W.~T. Roberts \emph{et~al.}, ``Laser power beaming for lunar night and permanently shadowed regions,'' \emph{Jet Propulsion}, vol. 818, pp. 354--1566, 2021.

\bibitem{microSat2Earth}
W.~C. Brown, ``Power/energy: Solar power satellites: Microwaves deliver the power: Kilometer-diameter antennas would be used in this super system to transmit power to ground-based rectifying antennas,'' \emph{IEEE Spectrum}, vol.~16, no.~6, pp. 36--43, 1979.

\bibitem{chen_novel_2019}
J.~Chen, L.~Yang, W.~Wang, H.-C. Yang, Y.~Liu, M.~O. Hasna, and M.-S. Alouini, ``A novel energy harvesting scheme for mixed {FSO}-{RF} relaying systems,'' \emph{{IEEE} Transactions on Vehicular Technology}, vol.~68, no.~8, pp. 8259--8263.

\bibitem{he_analysis_2021}
T.~He, L.~Zhang, G.~Zheng, C.~Yang, M.~Wang, and G.~Pan, ``Analysis and experiment of the laser wireless energy transmission efficiency based on the receiver of powersphere,'' \emph{{IEEE} Access}, vol.~9, pp. 55\,340--55\,351.

\bibitem{donmez_mitigation_2023}
B.~Donmez, I.~Azam, and K.~G. Karabulut, ``Mitigation of misalignment errors over inter-satellite {FSO} energy harvesting.''

\bibitem{alouiniEH2}
M.~Qaraqe, M.~Usman, A.~Serbes, I.~S. Ansari, and M.-S. Alouini, ``Power hotspots in space: Powering cubesats via inter-satellite optical wireless power transfer,'' \emph{IEEE Internet of Things Magazine}, vol.~5, no.~3, pp. 180--185, 2022.

\bibitem{stubbs2007impact}
T.~J. Stubbs, R.~R. Vondrak, and W.~M. Farrell, ``Impact of dust on lunar exploration,'' \emph{Dust in Planetary Systems}, vol. 643, pp. 239--243, 2007.

\bibitem{popel_dusty_2015}
S.~I. Popel, L.~M. Zelenyi, and B.~Atamaniuk, ``Dusty plasma sheath-like structure in the region of lunar terminator,'' \emph{Physics of Plasmas}, vol.~22, no.~12, p. 123701.

\bibitem{popel_dusty_2013}
S.~I. Popel, S.~I. Kopnin, A.~P. Golub’, G.~G. Dol’nikov, A.~V. Zakharov, L.~M. Zelenyi, and Y.~N. Izvekova, ``Dusty plasma at the surface of the moon,'' \emph{Solar System Research}, vol.~47, no.~6, pp. 419--429.

\bibitem{jin_wireless_2019}
K.~Jin and W.~Zhou, ``Wireless laser power transmission: A review of recent progress,'' \emph{{IEEE} Transactions on Power Electronics}, vol.~34, no.~4, pp. 3842--3859.

\bibitem{kim_multidisciplinary_2020}
K.-J. Kim and K.-H. Yu, ``Multidisciplinary design optimization for a solar-powered exploration rover considering the restricted power requirement,'' \emph{Energies}, vol.~13, no.~24, p. 6652, number: 24.

\bibitem{adaptivebeamdivergence}
R.~Harada, N.~Shibata, S.~Kaneko, T.~Imai, J.-I. Kani, and T.~Yoshida, ``Adaptive beam divergence for expanding range of link distance in {FSO} with moving nodes toward {6G},'' \emph{IEEE Photonics Technology Letters}, vol.~34, no.~20, pp. 1061--1064, 2022.

\bibitem{EHwindow}
G.~Pan, H.~Zhang, R.~Zhang, S.~Wang, J.~An, and M.-S. Alouini, ``Space simultaneous information and power transfer: An enhanced technology for miniaturized satellite systems,'' \emph{IEEE Wireless Communications}, vol.~30, no.~2, pp. 122--129, 2023.

\bibitem{ComprehensivePathLoss}
H.~Kotake, Y.~Abe, M.~Sekiguchi, T.~Fuse, H.~Tsuji, and M.~Toyoshima, ``Link budget design of adaptive optical satellite network for integrated non-terrestrial network,'' in \emph{IEEE International Conference on Space Optical Systems and Applications (ICSOS)}, 2022, pp. 240--247.

\bibitem{Shlomi_optimization_2004}
A.~Polishuk and S.~Arnon, ``Optimization of a laser satellite communication system with an optical preamplifier,'' \emph{J. Opt. Soc. Am. A}, vol.~21, no.~7, pp. 1307--1315, Jul. 2004.

\bibitem{craig_f_bohren_donald_r_huffman_absorption_1998}
{Craig F. Bohren, Donald R. Huffman}, ``Absorption and scattering by a sphere,'' in \emph{Absorption and Scattering of Light by Small Particles}.\hskip 1em plus 0.5em minus 0.4em\relax John Wiley \& Sons, Ltd, pp. 82--129.

\bibitem{liu_characterization_2008}
Y.~Liu, J.~Park, D.~Schnare, E.~Hill, and L.~A. Taylor, ``Characterization of lunar dust for toxicological studies. {II}: Texture and shape characteristics,'' \emph{Journal of Aerospace Engineering}, vol.~21, no.~4, pp. 272--279.

\bibitem{wriedt_mie_2012}
T.~Wriedt, ``Mie theory: A review,'' in \emph{The Mie Theory: Basics and Applications}, ser. Springer Series in Optical Sciences, W.~Hergert and T.~Wriedt, Eds.\hskip 1em plus 0.5em minus 0.4em\relax Springer, pp. 53--71.

\bibitem{yahia_haps_2022}
O.~B. Yahia, E.~Erdogan, G.~K. Karabulut, I.~Altunbas, and H.~Yanikomeroglu, ``{HAPS} selection for hybrid {RF}/{FSO} satellite networks,'' \emph{{IEEE} Transactions on Aerospace and Electronic Systems}, vol.~58, no.~4, pp. 2855--2867.

\bibitem{lacaze_gaps_2009}
B.~Lacaze, ``Gaps of free-space optics beams with the beer-lambert law,'' \emph{Applied Optics}, vol.~48, no.~14, p. 2702.

\bibitem{park_characterization_2008}
J.~Park, Y.~Liu, K.~D. Kihm, and L.~A. Taylor, ``Characterization of lunar dust for toxicological studies i: Particle size distribution,'' \emph{Journal of Aerospace Engineering}, vol.~21, no.~4, pp. 266--271.

\bibitem{vidwans_size_2022}
A.~Vidwans, S.~Choudhary, B.~Jolliff, J.~Gillis-Davis, and P.~Biswas, ``Size and charge distribution characteristics of fine and ultrafine particles in simulated lunar dust: Relevance to lunar missions and exploration,'' \emph{Planetary and Space Science}, vol. 210, p. 105392.

\bibitem{popel2018lunar}
S.~I. Popel, L.~M. Zelenyi, A.~Y. Dubinskii \emph{et~al.}, ``Lunar dust and dusty plasmas: Recent developments, advances, and unsolved problems,'' \emph{Planetary and Space Science}, vol. 156, pp. 71--84, 2018.

\bibitem{refractiveindex1}
A.~Fearnside, P.~Masding, and C.~Hooker, ``Polarimetry of moonlight: A new method for determining the refractive index of the lunar regolith,'' \emph{Icarus}, vol. 268, pp. 156--171.

\bibitem{refractiveindex2}
J.~D. Goguen, T.~C. Stone, H.~H. Kieffer, and B.~J. Buratti, ``A new look at photometry of the moon,'' \emph{Icarus}, vol. 208, no.~2, pp. 548--557.

\bibitem{draine_optical_1984}
B.~T. Draine and H.~M. Lee, ``Optical properties of interstellar graphite and silicate grains,'' \emph{The Astrophysical Journal}, vol. 285, p.~89.

\bibitem{Solarcells}
A.~Luque and S.~Hegedus, \emph{\BIBforeignlanguage{eng}{Handbook of {P}hotovoltaic {S}cience and {E}ngineering}}, 2nd~ed.\hskip 1em plus 0.5em minus 0.4em\relax Chichester, West Sussex, U.K.: Wiley, 2011.

\bibitem{algora_beaming_2022}
C.~Algora, I.~García, M.~Delgado, R.~Peña, C.~Vázquez, M.~Hinojosa, and I.~Rey-Stolle, ``Beaming power: Photovoltaic laser power converters for power-by-light,'' \emph{Joule}, vol.~6, no.~2, pp. 340--368.

\bibitem{lasertypes}
H.~Kaushal and G.~Kaddoum, ``Optical communication in space: Challenges and mitigation techniques,'' \emph{IEEE Communications Surveys \& Tutorials}, vol.~19, no.~1, pp. 57--96, 2017.

\bibitem{laser1064}
H.-Y. Lin, M.-Y. Wu, J.~Tang, M.-J. Zhao, and D.~Sun, ``Compact {Nd}:{YVO4} laser 1087.5 nm pumped by an edge-emitting {LD},'' \emph{Optik}, vol. 185, pp. 414--417, 2019.

\bibitem{spotdiameter}
A.~G. Alkholidi and K.~S. Altowij, ``Free {Space} {Optical} {Communications} — {Theory} and {Practices},'' in \emph{Contemporary {Issues} in {Wireless} {Communications}}, M.~Khatib, Ed., Rijeka, 2014.

\bibitem{EHCE}
R.~Jomen, F.~Tanaka, T.~Akiba, M.~Ikeda, K.~Kiryu, M.~Matsushita, H.~Maenaka, P.~Dai, S.~Lu, and S.~Uchida, ``Conversion efficiencies of single-junction {III}–{V} solar cells based on {InGaP}, {GaAs}, {InGaAsP}, and {InGaAs} for laser wireless power transmission,'' \emph{Japanese Journal of Applied Physics}, vol.~57, no. 8S3, Jul. 2018.

\end{thebibliography}

\thebiography

\begin{biographywithpic}
{Mohamed Naqbi}{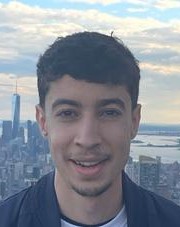}
is currently pursuing a double Master's degree in Aeronautical Engineering from École Polytechnique de Bruxelles and Aerospace Engineering from Polytechnique Montréal, located in Quebec, Canada. He obtained his Bachelor of Engineering degree with high distinction in Electromechanical Engineering from École Polytechnique de Bruxelles in 2021. His research interests are diverse and encompass energy harvesting in space, free-space optics, power beaming, optimization techniques, and machine learning applications.

\end{biographywithpic}

\begin{biographywithpic}
{Sébastien Loranger}{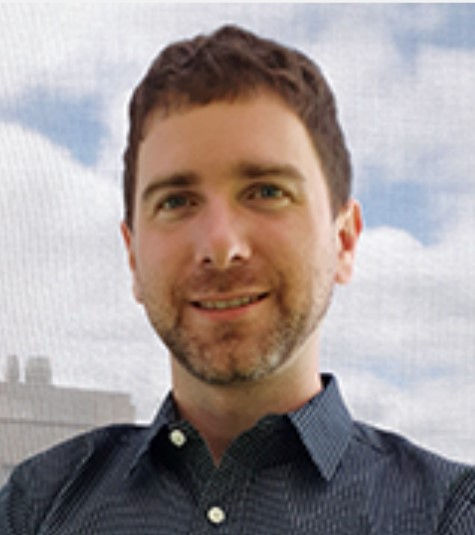}
is currently an assistant professor at Polytechnique Montreal, Montreal Qc, Canada. He received his
PhD at Polytechnique Montreal in Engineering Physics. He then did a postdoctoral fellowship at the renowned Max-Planck Institute in Erlangan, Germany, after which he worked as a R\&D researcher at Photonova Inc. and ITF Technologies before joining Polytechnique.
\end{biographywithpic} 

\begin{biographywithpic}
{Gunes Karabulut Kurt}{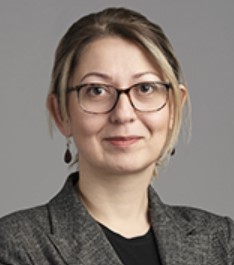}
is a Canada Research Chair (Tier 1) in New Frontiers in Space Communications and Associate Professor at Polytechnique Montréal,~Montréal,~QC,~Canada.~She is also an adjunct research professor at Carleton University.~Gunes  received the B.S. degree with high honors in electronics and electrical engineering from Bogazici University, Istanbul, Turkiye, in 2000 and the M.A.Sc.~and the Ph.D. degrees in electrical engineering from the University of Ottawa, ON, Canada, in 2002 and 2006, respectively. She worked in different technology companies in Canada and Turkiye, between 2005 and 2010. From 2010 to 2021, she was a professor at Istanbul Technical University. Gunes is a Marie Curie Fellow and has received the Turkish Academy of Sciences Outstanding Young Scientist (TÜBA-GEBIP) Award in 2019. She is serving as the secretary of IEEE Satellite and Space Communications Technical Committee,  the chair of the IEEE special interest group entitled “Satellite Mega-constellations: Communications and Networking” and also as an editor in 6 different IEEE journals. She is a member of the IEEE WCNC Steering Board and a Distinguished Lecturer of Vehicular Technology Society Class of 2022.  
\end{biographywithpic}

\end{document}